\documentclass{article}
\usepackage{iclr2027_conference,times}
\usepackage{amsmath,amsfonts,bm}

\def\eqref#1{equation~\ref{#1}}
\def\1{\bm{1}}

\DeclareMathAlphabet{\mathsfit}{\encodingdefault}{\sfdefault}{m}{sl}
\SetMathAlphabet{\mathsfit}{bold}{\encodingdefault}{\sfdefault}{bx}{n}

\usepackage{amsmath}
\usepackage{graphicx}
\usepackage{booktabs}
\usepackage{array}
\usepackage{hyperref}
\usepackage{url}
\usepackage{microtype}

\title{Evaluating Predictor-Corrector Dependence in Hard-Constrained Climate Emulators}

\newcommand{\AuthorSep}{\hspace{1.5em}}
\newcommand{\AuthorVSep}{\\[1.0ex]}

\newcommand{\Mila}{\textsuperscript{1}}
\newcommand{\McGill}{\textsuperscript{2}}
\newcommand{\UToronto}{\textsuperscript{3}}
\newcommand{\UMontreal}{\textsuperscript{4}}

\author{
    \parbox{\textwidth}{
        \vspace{3.5ex}
        \centering
        Tianyu Zhang\Mila\AuthorSep
        Xinyu Wang\McGill\AuthorSep
        Zhenghan Tai\UToronto\AuthorSep
        Xiao-Wen Chang\McGill\AuthorSep
        Yoshua Bengio\UMontreal
        \\[2.5ex]
        {
            \small\normalfont
            \Mila{}\,Mila -- Quebec Artificial Intelligence Institute\AuthorVSep
            \McGill{}\,McGill University\AuthorSep
            \UToronto{}\,University of Toronto\AuthorVSep
            \UMontreal{}\,Universit\'e de Montr\'eal
        }
    }
}
\iclrfinalcopy
\begin{document}
\maketitle

\begin{abstract}
Climate emulators often enforce physical constraints by correcting each state predicted by a neural network, but evaluating only the corrected forecast conflates predictor skill with the effect of the correction. We introduce an evaluation framework that separates these contributions by comparing corrected forecasts, pre-correction predictions along corrected trajectories, and autonomous rollouts with correction disabled. We further cross correction during training and inference to distinguish its immediate effect from how it shapes the learned predictor, and use matched nonphysical controls to test whether improvements are specific to the physical correction. In ACE2, correction substantially improves physical consistency. In a three-seed ACE2 pilot, models trained with correction produce less accurate uncorrected precipitation forecasts at short horizons than models trained without correction, but more accurate forecasts at 10 and 30 days. This contradicts our preregistered hypothesis that the deficit would grow with forecast horizon. In a separate nine-seed experiment, directly supervising the uncorrected output reverses its precipitation forecast deficit at 72 hours while maintaining the accuracy of corrected forecasts at 24 hours. In ConCerNet's heat-equation system, projection-trained models have higher uncorrected trajectory-based vector-field RMSE than models trained without projection, while restoring the physical projection recovers accuracy far more effectively than a matched nonphysical control. Together, these results show that hard corrections can reshape the predictors trained alongside them, creating dependencies that are not visible from corrected forecasts alone. Hard-constrained emulators should therefore be evaluated by separately measuring predictor accuracy, correction effects, and their interaction during training.
\end{abstract}

\section{Introduction}

Learned weather and climate models increasingly combine a neural predictor with a deterministic operation that enforces conservation, positivity, diagnostic closure, or prescribed boundary conditions \citep{beucler2021constraints,wattmeyer2025ace2}. These operations can improve a forecast while concealing errors in the state predicted by the network \citep{chapman2026hard}. Evaluation of the corrected output alone cannot distinguish an accurate learned transition from an inaccurate transition whose errors are subsequently removed.

Training may improve the predictor itself; the corrector may repair a largely unchanged prediction; or the predictor may adapt to a correction available at every step \citep{um2020solver,chapman2026hard}.

\begin{figure}[!htbp]
\centering
\includegraphics[width=\linewidth]{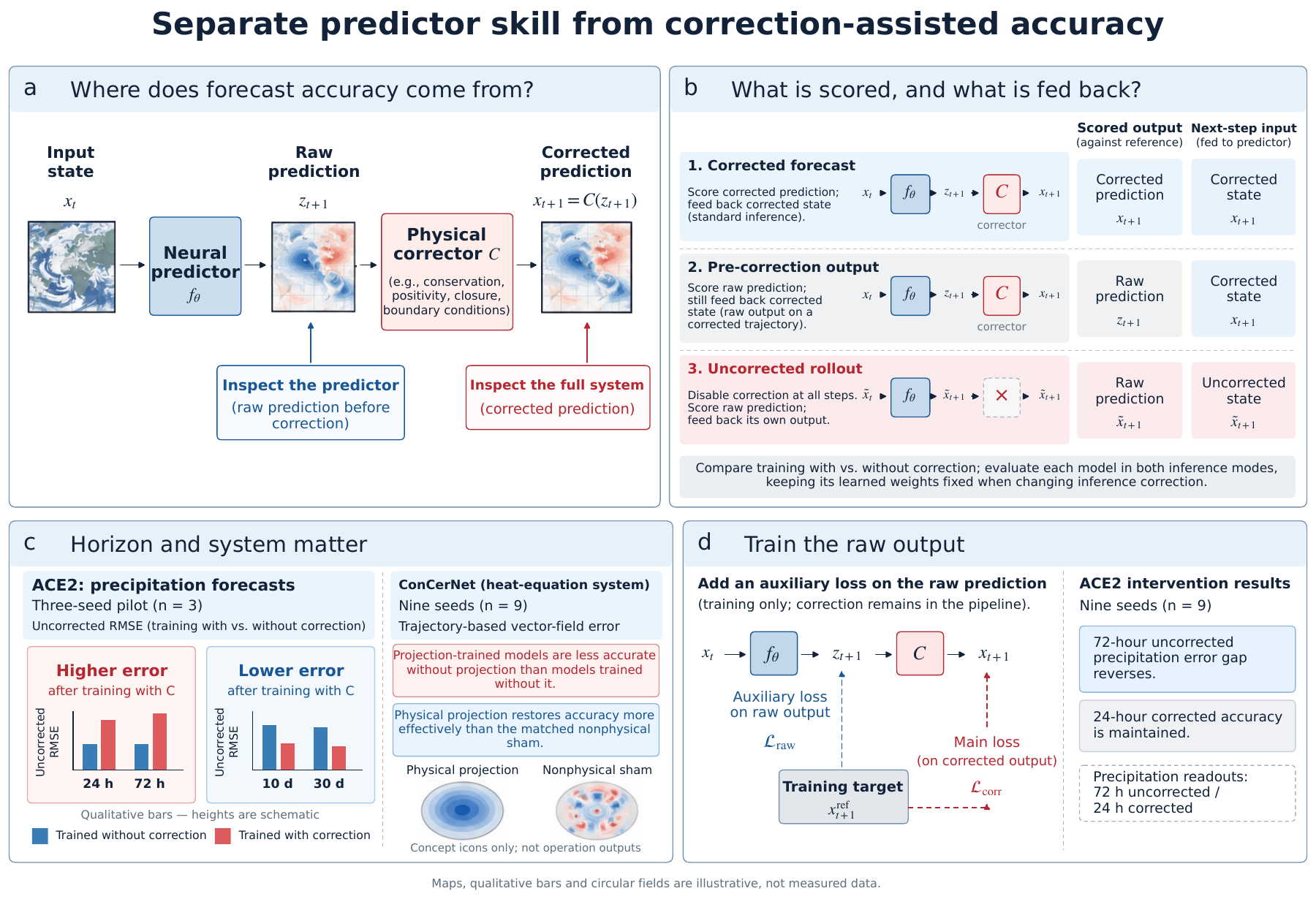}
\caption{\textbf{Separating predictor skill from correction-assisted accuracy.} (a) The neural predictor produces $z_{t+1}$ and the physical corrector $C$ produces $x_{t+1}$; other inputs are suppressed. (b) The scored output and recurrent input define three complementary readouts. Each trained model retains its weights across inference modes. (c) Schematic bars summarize the ACE2 pilot's horizon-dependent training effect; blue/red denote training without/with correction. ConCerNet compares uncorrected vector-field error and physical versus sham projection. Quantitative results appear in Figures~\ref{fig:c3-completed-contrasts} and~\ref{fig:concernet-decision}. (d) Dashed paths indicate training supervision. Table~\ref{tab:raw-loss-summary} reports the nine-seed intervention's 72-hour uncorrected and 24-hour corrected precipitation outcomes. The state maps and circular motifs are schematic.}
\label{fig:audit-overview}
\end{figure}

Our proposed evaluation uses three complementary comparisons. Matched controls test whether the physical correction contributes more than an equal computation budget or a nonphysical perturbation. We then evaluate network outputs immediately before correction along corrected trajectories. Finally, we run models trained with and without correction in both inference modes, keeping each model's learned weights fixed. Disabling correction throughout forecasting makes the predictor repeatedly receive its own uncorrected outputs, exposing the feedback-dependent input distributions studied in sequential prediction \citep{ross2011reduction}. Figure~\ref{fig:audit-overview} outlines evaluation (a,b), findings (c), and raw-output supervision (d).

ACE2 \citep{wattmeyer2025ace2} provides a revealing case study. Its corrector greatly improves conservation and positivity, but the precipitation advantage depends on forecast horizon. With correction disabled throughout forecasting, models trained with correction have higher precipitation RMSE than models trained without it at 24 and 72 h, but lower RMSE at 10 and 30 days. Direct supervision of the uncorrected output reduces the short-horizon deficit. ConCerNet's heat-equation system \citep{zhang2023concernet} exhibits a different pattern: projection-trained models rely strongly on the physical projection to recover predictive accuracy.

Our contribution is an experimental method for attributing forecast skill to a learned predictor, its corrector, and their interaction during training. We evaluate it through the ACE2 case study, a nine-seed ConCerNet experiment in a separate constrained dynamical system, and within-family comparisons of total-energy correction in ACE2S-SHiELD+.

% We also evaluate total-energy correction in ACE2S-SHiELD+, a model within the ACE2 family \citep{clark2026disentangling}.

\section{Related Work}

\textbf{Neural atmospheric models.} GraphCast and Pangu-Weather learn weather forecasts \citep{lam2023graphcast,bi2023pangu}; NeuralGCM couples learned components to differentiable dynamics \citep{kochkov2024neuralgcm}, and ACE2 targets weather-to-climate rollouts \citep{wattmeyer2025ace2}.

\textbf{Physics-guided and closed-loop learning.} Physical priors \citep{karniadakis2021physics} include analytic atmospheric constraints \citep{beucler2021constraints} and hard-constrained turbulence architectures \citep{mohan2023constraints}. Learned solver corrections \citep{um2020solver,dresdner2022spectral} and feedback-induced distribution shifts \citep{ross2011reduction} motivate evaluating how learned components behave when their outputs become future inputs.

\textbf{Corrector dependence and negative controls.} \citet{chapman2026hard} show that precipitation rescaling can hide deterioration in raw network outputs and study remedies using raw supervision and an imbalance penalty. We additionally remove correction from autoregressive feedback and cross training with inference conditions. Spatial shams provide target-independent negative controls \citep{lipsitch2010negative} by disrupting physical alignment.

\textbf{Constraint-enforcing operations.} Optimization layers \citep{amos2017optnet}, completion-and-correction \citep{donti2021dc3}, homeomorphic projections \citep{liang2023homeomorphic}, and constrained generation \citep{cheng2025gradientfree} incorporate feasibility into learned predictions. ConCerNet preserves invariants through vector-field projection \citep{zhang2023concernet}, providing a setting for our crossed training--inference evaluation.

\section{Method}

\subsection{Predictor--Corrector Formulation}\label{sec:readouts}

We first describe the state-correction formulation used for ACE2 \citep{wattmeyer2025ace2}. Let $x_t$ denote the model state and $b_t$ the prescribed boundary or forcing information. The predictor produces
\begin{equation}
z_{t+1}=f_\theta(x_t,b_t),
\end{equation}
and the physical corrector produces the next state,
\begin{equation}
x_{t+1}=C(z_{t+1},x_t,b_t).
\end{equation}
We evaluate prediction error against the reference state $x_{t+1}^{\mathrm{ref}}$ in three settings, distinguished by the scored output and the next-step input in Figure~\ref{fig:audit-overview}b.

\paragraph{Corrected rollout.}
We evaluate $x_{t+1}$ and feed it into the next prediction step. This measures the forecast accuracy of the predictor--corrector system.

\paragraph{Pre-correction prediction along a corrected rollout.}
Along the same corrected trajectory, we evaluate $z_{t+1}$ while feeding the corrected state $x_{t+1}$ into the next step. This measures the accuracy of the network output conditional on a history of corrected states.

\paragraph{Uncorrected rollout.}
We disable correction throughout forecasting and recursively feed the network's outputs back into the predictor:
\begin{equation}
\tilde{x}_{t+1}=f_\theta(\tilde{x}_t,b_t),
\qquad
\tilde{x}_0=x_0.
\end{equation}
We evaluate $\tilde{x}_{t+1}$ against $x_{t+1}^{\mathrm{ref}}$. This measures forecast accuracy without correction, including the effects of earlier predictions on subsequent inputs. For the same predictor, initial state, and prescribed forcing, $\tilde{x}_1=z_1$. Later predictions can differ because the two rollouts feed back different states.

\subsection{Matched-Control Evaluation in ACE2}

We compare the official physical correction with four controls: no correction, longitude shift, latitude-band permutation, and spectrum-preserving phase randomization. All models receive 2400 adaptation updates, with the common update budget selected using validation data. Each model uses the same correction condition during adaptation and forecasting, so these comparisons include both adaptation and inference effects.

Evaluation uses nine seeds, four months from 1940, forecast horizons of 24 and 72 hours, and three variables: near-surface temperature, 500 hPa height, and precipitation. The evaluation months are held out from local adaptation but fall within the released model's pretraining period \citep{wattmeyer2025ace2}.

The three spatial shams disrupt the spatial alignment of the correction. Spatial phase randomization follows the surrogate-data principle of preserving spectral properties while disrupting other structure \citep{theiler1992surrogate}. We evaluate predictive accuracy together with conservation, positivity, correction magnitude, and decorrelation.

\subsection{Training--Inference Evaluation in ACE2}

We independently vary correction during adaptation and forecasting. Models adapted with and without correction are each evaluated with correction enabled and disabled, yielding four combinations. Within each adapted model, learned weights, normalization, initial states, prescribed forcing, and evaluation settings remain fixed across inference conditions.

This pilot uses three seeds, four months held out from adaptation, and forecast horizons of 24 hours, 72 hours, 10 days, and 30 days.

For a given variable and forecast horizon, let $R_{a,i,s}$ denote the RMSE for seed $s$, averaged over the four evaluation months. The indices $a$ and $i$ indicate correction during adaptation and inference, respectively, with 1 denoting enabled and 0 denoting disabled. The relative benefit of inference correction is
\begin{equation}\label{eq:inference-benefit}
I_{a,s}
=
100\frac{R_{a,0,s}-R_{a,1,s}}{R_{a,0,s}}.
\end{equation}
We evaluate this effect separately for predictors adapted with and without correction. The adaptation comparison with inference correction disabled is
\begin{equation}\label{eq:adaptation-benefit}
A_s
=
100\frac{R_{0,0,s}-R_{1,0,s}}{R_{0,0,s}}.
\end{equation}
Positive $A_s$ indicates lower uncorrected RMSE after adaptation with correction. We also report $I_{1,s}-I_{0,s}$ in percentage points. This contrast measures the difference between proportional inference benefits, each normalized by the corresponding predictor's uncorrected RMSE.

Our preregistered hypothesis predicts that adaptation with correction reduces uncorrected forecast accuracy relative to adaptation without correction, with the deficit increasing across forecast horizons.

\subsection{Auxiliary Supervision of Pre-Correction Predictions}

Motivated by the raw-supervision remedy studied by \citet{chapman2026hard}, we supervise pre-correction precipitation while retaining the official corrector during adaptation. The per-step objective is
\begin{equation}
\label{eq:raw-supervision}
\mathcal{L}_{\mathrm{step}}=\mathcal{L}_{\mathrm{main}}+\lambda\,\operatorname{MSE}\!\left(N_p(z_{t+1,p}),N_p(x^{\mathrm{ref}}_{t+1,p})\right),
\end{equation}
where $p$ denotes surface precipitation and $N_p$ is its normalization in the main loss. The auxiliary term averages squared errors over batch and spatial entries, supervises only precipitation, and leaves the main loss unchanged.

We select $\lambda$ from $\{0.005,0.02,0.1\}$ using March 1940 validation forecasts averaged over seeds 7, 31, and 71. Eligible candidates have corrected 72-hour precipitation RMSE no more than 1\% above the original model. We select the candidate with the lowest uncorrected 72-hour RMSE, breaking ties toward the smaller weight. This selects $\lambda=0.005$ before held-out evaluation. All models receive 2400 adaptation updates. We evaluate the selected model with correction enabled and disabled; Appendix~\ref{app:supervision-method} gives the evaluation design.

\subsection{Evaluation in a Heat-Equation System}

We apply the crossed training--inference design to ConCerNet's 101-dimensional heat-equation system \citep{zhang2023concernet}. Our experiment uses a known analytic invariant in place of learning one. The network predicts a vector field, and physical projection subtracts the mean of its components from each component. The projected vector field therefore sums to zero, enforcing conservation of the sum of state components at the vector-field level.

Networks trained with and without projection start from identical weights and use the same training data and minibatch order. Each network is evaluated with and without physical projection. An additional sham projection is matched in norm and constructed independently of evaluation targets.

The prediction metric is trajectory-based vector-field RMSE. At a forecast horizon of one time unit, we evaluate the model's vector field at the forecast state and the reference vector field at the reference state, then compute their RMSE. This metric combines differences in the vector fields and their evaluation states; state RMSE is a separate quantity.

\subsection{Statistical Analysis}

\paragraph{Aggregation before comparison.}
In the matched-control experiment, RMSE is averaged over seven paired forecast starts within each seed and month. Relative effects are then computed within month and averaged equally over months and seeds. The training--inference pilot first averages RMSE over four months within seed, then computes paired relative effects. Thus, the former averages month-specific relative effects, whereas the latter compares month-averaged errors. Appendix~\ref{app:error-aggregation} gives the estimands.

\paragraph{ACE2 matched-control evaluation.}
Training seed is the independent unit \citep{bouthillier2021variance}; months and starts are nested paired observations. Across 24 confirmatory comparisons (four controls, two horizons, three variables), we report seed-level means, hierarchical bootstrap 95\% intervals \citep{efron1979bootstrap}, and exact one-sided sign-flip tests \citep{ernst2004permutation}, with Holm correction \citep{holm1979multiple}. Long-horizon analyses are descriptive.

\paragraph{ACE2 training-inference evaluation.}
For the three-seed pilot, Figure~\ref{fig:c3-completed-contrasts} shows mean paired effects and descriptive 95\% Student-$t$ intervals, $\bar{e}\pm t_{0.975,2}s_e/\sqrt{3}$, where $s_e$ is the sample standard deviation across seeds. Table~\ref{tab:c3-factorial-contrasts} reports the same means and observed seed ranges, not confidence intervals. With three seeds, normality cannot be assessed reliably, and the minimum one-sided exact sign-test $p$-value is 0.125; these analyses remain descriptive.

\paragraph{ACE2 direct-supervision evaluation.}
The direct-supervision analysis uses nine adaptation seeds. We use bootstrap resampling \citep{efron1979bootstrap}, with adaptation seeds as clusters: 20,000 resamples preserve paired model comparisons within each resampled seed.

\paragraph{ConCerNet evaluation.}
We evaluate nine independent seeds at 0.1, 0.5, 1.0, and 2.0 model-time units using 32 held-out initial states. The confirmatory horizon is 1.0; other horizons are descriptive. Intervals use 20,000 paired seed-bootstrap resamples. Lower confidence bounds are the lower endpoints of two-sided 95\% percentile intervals. Table~\ref{tab:concernet-decision} lists invariant-drift, prediction-error, physical--sham, and seed-consistency criteria.
\section{Results}

\subsection{Practical Gains in Temperature}

Across the 24 matched-regime comparisons, 23 favor the official correction and remain significant after Holm correction \citep{holm1979multiple}. Eight mean effects meet the practical-importance thresholds, all for temperature. The 72-hour precipitation comparison against no correction is inconclusive. Table~\ref{tab:c2-fixed-contrasts} reports all effects and their uncertainty; Figure~\ref{fig:c2-primary-effects} in Appendix~\ref{app:matched-results} provides the corresponding interval plot.

\begin{table}[!htb]
\caption{\textbf{ACE2 matched-control forecast improvements.} Relative RMSE reduction (\%) with official correction versus each control; correction conditions differ during both adaptation and inference.}
\label{tab:c2-fixed-contrasts}
\centering
\fontsize{9}{10}\selectfont
\setlength{\tabcolsep}{3pt}
\renewcommand{\arraystretch}{1.15}
\begin{tabular*}{\linewidth}{@{\extracolsep{\fill}}l*{6}{c}@{}}
\toprule

& \multicolumn{2}{c}{2 m temperature} & \multicolumn{2}{c}{500 hPa height} & \multicolumn{2}{c}{Precipitation} \\
\cmidrule(lr){2-3}\cmidrule(lr){4-5}\cmidrule(l){6-7}
Control & 24 h & 72 h & 24 h & 72 h & 24 h & 72 h \\
\midrule
No correction
& \shortstack{\textbf{0.98}$^{\dagger}$\\{\footnotesize [0.90, 1.06]}}
& \shortstack{\textbf{0.68}$^{\dagger}$\\{\footnotesize [0.44, 0.92]}}
& \shortstack{4.80$^{\dagger}$\\{\footnotesize [3.97, 5.66]}}
& \shortstack{0.87$^{\dagger}$\\{\footnotesize [0.48, 1.27]}}
& \shortstack{2.62$^{\dagger}$\\{\footnotesize [2.26, 2.98]}}
& \shortstack{0.00\\{\footnotesize [-0.53, 0.54]}} \\[3pt]
Longitude shift
& \shortstack{\textbf{0.95}$^{\dagger}$\\{\footnotesize [0.89, 1.01]}}
& \shortstack{\textbf{0.83}$^{\dagger}$\\{\footnotesize [0.64, 1.02]}}
& \shortstack{0.92$^{\dagger}$\\{\footnotesize [0.78, 1.06]}}
& \shortstack{0.69$^{\dagger}$\\{\footnotesize [0.50, 0.89]}}
& \shortstack{3.53$^{\dagger}$\\{\footnotesize [3.15, 3.90]}}
& \shortstack{1.79$^{\dagger}$\\{\footnotesize [1.34, 2.23]}} \\[3pt]
\shortstack[l]{Latitude-band\\permutation}
& \shortstack{\textbf{0.95}$^{\dagger}$\\{\footnotesize [0.89, 1.01]}}
& \shortstack{\textbf{0.83}$^{\dagger}$\\{\footnotesize [0.63, 1.03]}}
& \shortstack{0.92$^{\dagger}$\\{\footnotesize [0.79, 1.06]}}
& \shortstack{0.70$^{\dagger}$\\{\footnotesize [0.50, 0.89]}}
& \shortstack{3.52$^{\dagger}$\\{\footnotesize [3.13, 3.87]}}
& \shortstack{1.80$^{\dagger}$\\{\footnotesize [1.36, 2.24]}} \\[3pt]
Phase randomization
& \shortstack{\textbf{0.95}$^{\dagger}$\\{\footnotesize [0.89, 1.01]}}
& \shortstack{\textbf{0.85}$^{\dagger}$\\{\footnotesize [0.65, 1.04]}}
& \shortstack{0.97$^{\dagger}$\\{\footnotesize [0.84, 1.11]}}
& \shortstack{0.74$^{\dagger}$\\{\footnotesize [0.54, 0.93]}}
& \shortstack{3.60$^{\dagger}$\\{\footnotesize [3.21, 3.98]}}
& \shortstack{1.86$^{\dagger}$\\{\footnotesize [1.40, 2.31]}} \\
\bottomrule
\end{tabular*}
\par\smallskip
\begin{minipage}{\linewidth}
\footnotesize
Positive values favor official correction. Brackets: pointwise hierarchical bootstrap 95\% intervals over nine seeds, four months per seed, and seven paired starts per month.
$\dagger$: Holm-adjusted one-sided sign-flip $p=0.047$ across 24 tests; the unmarked comparison has $p=0.514$.
\textbf{Bold}: the unrounded mean meets the practical threshold (temperature: 0.5\%; height and precipitation: 5\%).
The 72-hour precipitation mean versus no correction is slightly negative before rounding to 0.00\%.
\end{minipage}
\end{table}

\subsection{Physical Consistency and Long-Horizon Error}

A separate three-seed pilot evaluates ACE2's conservation and nonnegativity corrector \citep{wattmeyer2025ace2} using seven October 1940 forecast starts (Appendix~\ref{app:physical-pilot}). Mean maximum dry-air drift was 0.0625 Pa with official correction during both adaptation and forecasting, compared with 1332 Pa without correction at either stage. Mean water-budget residual magnitude was $1.15 \times 10^{-12}$ versus $7.93 \times 10^{-5}$ kg m$^{-2}$ s$^{-1}$, and the frequency of negative precipitation was 0\% versus 33.1\%.

In the longer matched-regime evaluations, precipitation RMSE was higher with correction at both stages than without correction at either stage: by 2.46\% over the 10-day rollout window and 1.75\% over the 30-day window. These effects are based on RMSE averaged over each window. No divergence was observed. Figure~\ref{fig:c2-long-horizon} reports the variable-specific effects.

\subsection{Horizon-Dependent Reversal of the Uncorrected Gap}

\begin{figure}[!htbp]
\centering
\includegraphics[width=\linewidth]{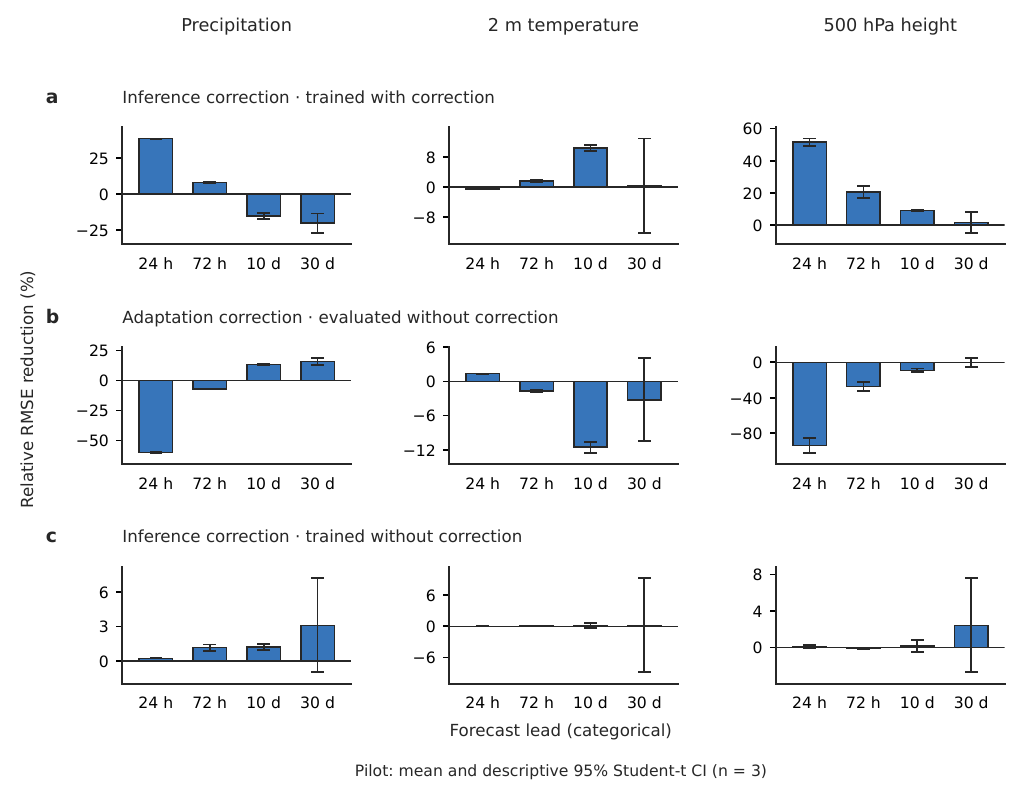}
\caption{\textbf{ACE2 correction effects in the three-seed pilot.} Rows show the effect of enabling inference correction after adaptation with correction (a), adaptation with versus without correction under uncorrected inference (b), and enabling inference correction after adaptation without correction (c). Bars show mean paired relative RMSE reductions; capped error bars show descriptive 95\% Student-$t$ intervals ($n=3$, two degrees of freedom). Positive values favor inference correction in a/c and adaptation with correction in b. Within each seed, RMSE is averaged over four months before computing relative effects. Forecast leads are categorical and vertical scales differ across panels.}
\label{fig:c3-completed-contrasts}
\end{figure}

With correction disabled throughout forecasting, predictors adapted with correction have higher precipitation RMSE at short horizons and lower RMSE at longer horizons than predictors adapted without correction (Figure~\ref{fig:c3-completed-contrasts}; Table~\ref{tab:c3-factorial-contrasts}). Mean seed-level relative increases are 59.7\% at 24 hours and 7.13\% at 72 hours; the 72-hour increases range from 7.04\% to 7.17\% across the three seeds. At 10 and 30 days, the comparison reverses in all three seeds, with mean relative RMSE reductions of 13.48\% and 15.69\%. This reversal contradicts our preregistered hypothesis that the uncorrected deficit grows with forecast horizon.

The pattern also varies across variables. At 24 hours, adaptation with correction increases uncorrected 500 hPa height RMSE by 93.72\% on average, while enabling inference correction for that predictor reduces its height RMSE by 51.58\%. At 10 days, the 13.48\% reduction in uncorrected precipitation RMSE contrasts with increases of 11.51\% in temperature RMSE and 9.17\% in height RMSE. The improvement at 10 days is therefore specific to precipitation among the three evaluated variables.

\subsection{Direct Supervision Reduces the Short-Horizon Gap}

Direct supervision reverses the uncorrected 72-hour precipitation RMSE gap in all nine seeds and lowers corrected 24-hour RMSE on average (Table~\ref{tab:raw-loss-summary}). The mean uncorrected gap changes from +4.92\% to -4.43\% relative to the same-seed predictor adapted without correction. This experiment evaluates June and September 1940, whereas the training--inference pilot averages four months; the baseline gaps therefore refer to different evaluation sets. Figure~\ref{fig:raw-loss-decision} shows the paired seed-level changes.

\begin{table}[!htb]
\caption{\textbf{Direct-supervision effects on precipitation forecasts ($n=9$).}}
\label{tab:raw-loss-summary}
\centering
\begin{tabular*}{\linewidth}{@{\extracolsep{\fill}}lrr@{}}
\toprule
Comparison & Mean & 95\% interval \\
\midrule
\multicolumn{3}{@{}l}{\textbf{A. Uncorrected forecasts: 72-hour precipitation}} \\
Original model: gap (\%) & $+4.92$ & $[+4.87,\,+4.97]$ \\
Direct supervision: gap (\%) & $-4.43$ & $[-4.47,\,-4.39]$ \\
Paired gap change: supervised $-$ original (pp) & $-9.35$ & $[-9.39,\,-9.32]$ \\
\addlinespace[2pt]
\multicolumn{3}{@{}l}{\textbf{B. Corrected forecasts: 24-hour precipitation}} \\
Direct supervision: RMSE change (\%) & $-0.67$ & $[-0.73,\,-0.59]$ \\
\bottomrule
\end{tabular*}
\par\smallskip
\begin{minipage}{\linewidth}
\footnotesize
References: A, same-seed model adapted without correction; B, original correction-adapted model. RMSE averages June and September 1940 within seed. Intervals: 20,000 paired seed-bootstrap resamples; gap changes are paired within seed (pp: percentage points). All 9/9 seeds reverse the gap and improve corrected accuracy. The corrected upper bound ($-0.59\%$) is below the prespecified $+1\%$ allowable increase.
\end{minipage}
\end{table}

\begin{figure}[!htbp]
\centering
\includegraphics[width=\linewidth]{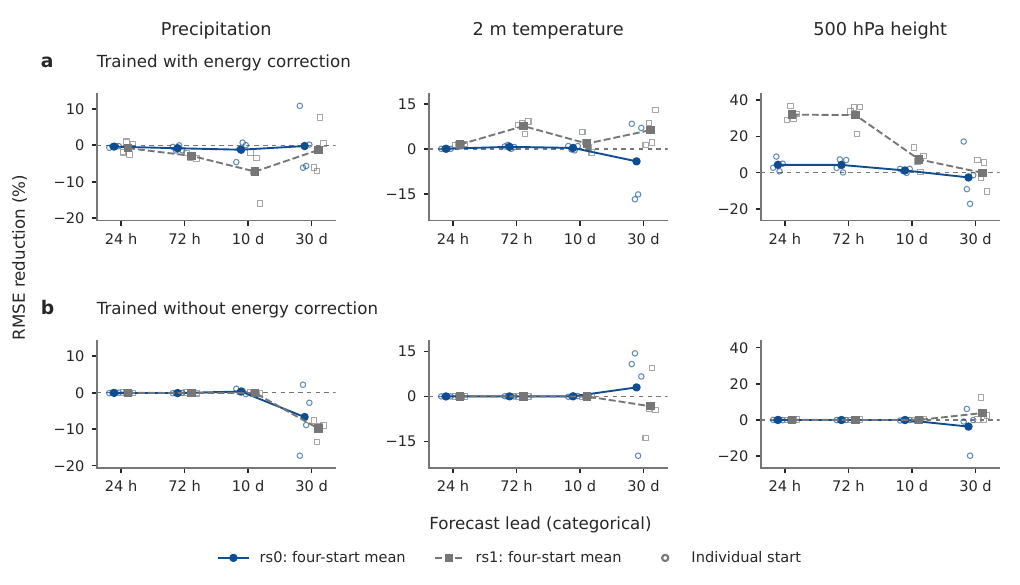}
\caption{\textbf{ACE2S energy-correction effects at fixed checkpoints.} Rows enable inference correction after adaptation with (a) or without (b) it; positive RMSE reductions favor correction. Blue circles/solid lines denote rs0; gray squares/dashed lines denote rs1. Hollow marks show January, April, July and October 1979 starts, ordered left to right within each seed--lead group; filled marks show their mean. Forecast starts are nested within checkpoint families. Leads are categorical; variable-specific scales match across rows.}
\label{fig:ace2s-replication}
\end{figure}
\begin{figure}[!htbp]
\centering
\includegraphics[width=\linewidth]{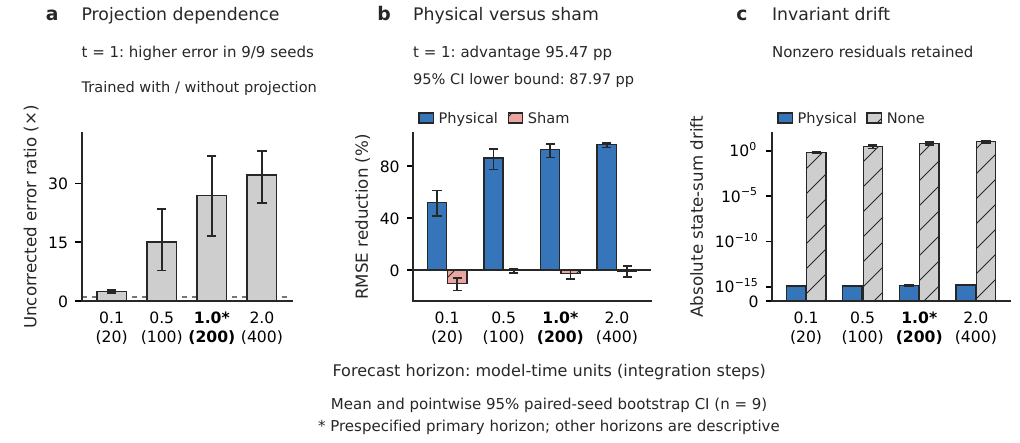}
\caption{\textbf{ConCerNet projection dependence across forecast horizons.} The horizontal axis is elapsed time from initialization in model-time units, not hours or days; parentheses give integration steps ($\Delta t=0.005$). Groups are categorical, not linearly spaced in time. The starred $t=1$ (200 steps) is the prespecified primary horizon; others are descriptive. Bars: nine-seed means; error bars: pointwise 95\% CIs from 20,000 paired seed-bootstrap resamples; 32 initial conditions per seed. (a) Uncorrected RMSE ratio, trained with/without projection; dashed line: equal error. (b) Physical (blue) and sham (pink, hatched) RMSE reductions versus no projection. Annotation: paired physical--sham difference. (c) Physical (blue) and no-projection (gray, hatched) state-sum drift. Panels b--c use projection-trained models; errors are trajectory-based vector-field RMSE, not state RMSE. Drift bars start at zero; the axis is linear below $10^{-16}$ and logarithmic above. Statistics use untransformed values; nonzero residuals are retained.}
\label{fig:concernet-decision}
\end{figure}

\subsection{Within-Family Energy-Correction Evaluation}

At fixed ACE2S-SHiELD+ checkpoints \citep{clark2026disentangling}, enabling energy correction reduces 24-hour 500 hPa height RMSE by 4.29\% and 32.00\% in the two checkpoint families trained with correction, rs0 and rs1 (Figure~\ref{fig:ace2s-replication}). For checkpoints trained without energy correction, the corresponding 24-hour height effects are below 0.1\%, while enabling correction increases 30-day precipitation RMSE by 6.68\% and 9.85\%. Each effect averages four initialization dates. These descriptive comparisons show variation across variables, horizons, and checkpoints.

These fixed-checkpoint comparisons hold learned weights constant when toggling inference correction. A separate comparison evaluates checkpoints trained with and without energy correction, with correction disabled at inference (Figure~\ref{fig:ace2s-budget-context}). The released rs1 checkpoints used 213408 training batches with energy correction and 328320 without it. That between-checkpoint comparison therefore includes both training-budget and correction-condition differences; the fixed-checkpoint results above isolate the inference change.

\subsection{Projection Dependence in ConCerNet}

In our analytic-invariant adaptation of ConCerNet \citep{zhang2023concernet}, the mean seed-level ratio of uncorrected vector-field error for projection-trained models to that for models trained without projection is 26.87 at one model-time unit. Applying physical projection reduces the projection-trained models' error by 92.65\% on average, whereas the norm-matched sham increases it by approximately 2.82\%. Both changes use the same model's error without inference projection as the reference.

Figure~\ref{fig:concernet-decision} extends these comparisons across four forecast times with pointwise 95\% seed-bootstrap intervals. At the primary time $t=1$, the physical--sham error-reduction difference is 95.47 percentage points. Table~\ref{tab:concernet-decision} reports the prespecified criteria and lower confidence bounds using the excess error ratio (total ratio minus one).

\section{Discussion}

In the ACE2 pilot, the effect of correction varies with adaptation history and forecast horizon. At longer horizons, the predictor adapted with correction has lower uncorrected precipitation RMSE than the predictor adapted without it, yet enabling its corrector increases RMSE. For the predictor adapted without correction, enabling correction reduces mean precipitation RMSE at every evaluated horizon. These results motivate evaluating coupled trajectories, as in prognostic parameterization tests \citep{brenowitz2018prognostic} and solver-in-the-loop learning \citep{um2020solver}.

The variable-specific patterns in Table~\ref{tab:c3-factorial-contrasts} suggest useful diagnostics. Following ACE2's separate assessments of bias and variability \citep{wattmeyer2025ace2}, we propose examining precipitation bias, variability, and spatial agreement alongside errors in other fields. Tracking correction magnitudes and water-budget residuals over forecast lead could help assess whether changes in the applied correction coincide with the reversal.

The direct-supervision experiment links the short-horizon deficit to the adaptation objective: supervising pre-correction precipitation reverses the 72-hour uncorrected RMSE gap while also improving corrected 24-hour accuracy (Table~\ref{tab:raw-loss-summary}). This complements the raw-amplitude stabilization studied by \citet{chapman2026hard} by evaluating the intervention under recursive forecasting.

ConCerNet extends predictor-corrector dependence to a constrained dynamical system with an analytic invariant \citep{zhang2023concernet}. Physical projection recovers prediction accuracy that a norm-matched sham does not (Figure~\ref{fig:concernet-decision}). Together, the ACE2 and ConCerNet results demonstrate the value of evaluating predictor accuracy and correction effects separately, including under crossed training--inference conditions.

\section{Limitations}

The climate experiments use adapted ACE2-family checkpoints with prescribed sea-surface temperature and sea ice \citep{wattmeyer2025ace2}. Their 1940 evaluation months fall within checkpoint pretraining, leaving generalization to unseen periods untested. The horizon-dependent reversal is based on three seeds, while the longer-horizon matched-regime comparisons remain descriptive. Direct supervision was evaluated at 72 hours without correction and 24 hours with correction. Extending these findings beyond the ACE2 family and the heat-equation ConCerNet system to other models and correction schemes remains an open direction.

\section{Conclusion}

Separating corrected forecasts, pre-correction outputs, and uncorrected rollouts reveals how physical correction interacts with the learned predictor. Across the tested ACE2 and ConCerNet settings, this interaction depends on training conditions, forecast horizon, and system. Direct supervision improves the short-horizon ACE2 predictor while preserving corrected accuracy. Evaluating these components separately therefore provides a clearer account of forecast skill in models that use physical correction.

% \section{Broader Impact}

% Component-level evaluation can help researchers diagnose forecast failures and assess changes to physically constrained models. Reporting both physical validity and predictive error makes the trade-offs of a correction visible, supporting more transparent decisions about model deployment and further development.

\newpage
\section*{AI use statement}

We used generative AI tools to assist with coding, experiment management, manuscript editing, and draft figure illustrations. The authors reviewed the resulting code, experimental settings, text, and figures and take responsibility for the final manuscript and accompanying artifacts.

% \section*{Reproducibility statement}

% The accompanying material provides prespecified designs, source tables, per-seed results, and verification records organized by experiment. The appendix identifies the source records and reproduction entry points; the technical documentation specifies execution order and environment requirements.

\bibliography{iclr2027_conference}
\bibliographystyle{iclr2027_conference}

\clearpage
\appendix
\section{Evaluation Definitions}

\subsection{Prediction and Correction}

We index the corrector in Section~\ref{sec:readouts} by condition $d$, writing $C_d$. With correction disabled, $C_0(z_{t+1},x_t,b_t)=z_{t+1}$. The three readouts and their feedback paths are defined in the main text and Figure~\ref{fig:audit-overview}b. Raw outputs are reconstructed by subtracting the recorded correction from the corrected output at each step of the corresponding rollout.

\subsection{Evaluation Contrasts}

Table~\ref{tab:app-estimands} summarizes the four training--inference contrasts. Training includes local adaptation of released checkpoints.

\begin{table}[!htbp]
\caption{Training--inference evaluation contrasts. Correction is enabled or disabled at every forecast step.}
\label{tab:app-estimands}
\centering
\small
\setlength{\tabcolsep}{4pt}
\renewcommand{\arraystretch}{1.15}
\begin{tabular}{@{}
  >{\raggedright\arraybackslash}p{\dimexpr0.30\linewidth-\tabcolsep\relax}
  >{\raggedright\arraybackslash}p{\dimexpr0.25\linewidth-2\tabcolsep\relax}
  >{\raggedright\arraybackslash}p{\dimexpr0.45\linewidth-\tabcolsep\relax}
@{}}
\toprule
\textbf{Contrast} & \textbf{Held fixed} & \textbf{Comparison and interpretation} \\
\midrule
Inference correction after training with correction
& Learned weights of the predictor trained with correction.
& Enable versus disable correction during forecasting; measures its effect on the same predictor. \\
\addlinespace
Training comparison without inference correction
& Correction disabled throughout both forecasts.
& Compare predictors trained with versus without correction; measures the difference between their uncorrected forecasts. \\
\addlinespace
Inference correction after training without correction
& Learned weights of the predictor trained without correction.
& Enable versus disable correction during forecasting; measures the effect of adding correction without retraining. \\
\addlinespace
Difference between inference effects
& Weights within each inference pair; metric, horizon, effect scale, and sign convention.
& Subtract the third-row inference effect from the first-row effect; measures the difference in relative correction benefits between the two predictors. \\
\bottomrule
\end{tabular}
\end{table}

We use the seed-level RMSE $R_{a,i,s}$ and effects $I_{a,s}$ and $A_s$ defined in Equations~\ref{eq:inference-benefit}--\ref{eq:adaptation-benefit}.
The difference between inference effects is
\begin{equation}
J_s=I_{1,s}-I_{0,s},
\end{equation}
reported in percentage points. Each inference effect is normalized by its own predictor's uncorrected RMSE. Thus, $J_s$ describes a difference between proportional improvements. We compute the contrast within each seed before summarizing its mean and range.

\subsection{Aggregation of Prediction Errors}
\label{app:error-aggregation}

The following expressions formalize the aggregation procedures described in the main text.

For the matched-control experiment, let $\overline{R}_{d,s,m}$ denote mean RMSE over seven paired forecast starts for correction condition $d$, seed $s$, and month $m$. The improvement relative to control $d$ is
\begin{equation}
G_{d,s,m}
=
100\frac{
\overline{R}_{d,s,m}-\overline{R}_{\mathrm{official},s,m}
}{
\overline{R}_{d,s,m}
}.
\end{equation}
These relative effects are averaged equally over four months and then over nine training seeds.

The training--inference effects use Equations~\ref{eq:inference-benefit}--\ref{eq:adaptation-benefit}.

The long-horizon matched-regime experiment uses window-averaged error. For each seed, month, and correction condition, spatial RMSE uses cosine-latitude weighting, as in WeatherBench \citep{rasp2020weatherbench}, and is averaged over all saved six-hour forecast leads and initialization dates within the rollout window. The relative change from the matched no-correction model is
\begin{equation}
H_{d,s,m}
=
100\left(\frac{R_{d,s,m}}{R_{0,s,m}}-1\right).
\end{equation}
Positive $H_{d,s,m}$ indicates higher RMSE with correction condition $d$.

\section{Experimental and Statistical Details}

\subsection{Direct Supervision of Uncorrected Outputs}
\label{app:supervision-method}

Equation~\ref{eq:raw-supervision} defines the auxiliary objective; its normalization and validation-only weight selection are described in the main text. Here we give the prespecified initiation rule and held-out evaluation design.

The preregistered joint hypothesis requires a 72-hour uncorrected precipitation deficit in all three pilot seeds, growth of the deficit across horizons, and a mean 24-hour precipitation RMSE reduction of at least 0.5\% when correction is restored. The direct-supervision experiment is initiated when the first and third criteria hold.

The direct-supervision evaluation uses nine adaptation seeds (7, 11, 23, 31, 43, 59, 71, 89, and 101) and two evaluation months, June and September 1940. Forecasts are evaluated at 24, 72, 240, and 720 hours. The supervised model is run with and without correction and compared with three matched baselines: the correction-trained model with correction enabled or disabled, and the model trained and evaluated without correction. The reported intervention criteria use the 72-hour uncorrected precipitation gap and the 24-hour corrected precipitation RMSE change.

\subsection{Statistical Analysis}

To account for training randomness \citep{bouthillier2021variance}, training seed is the independent unit. Months and forecast starts are paired observations nested within seed.

For the matched-control experiment, we use bootstrap resampling \citep{efron1979bootstrap} to report seed-level effects and pointwise 95\% hierarchical percentile-bootstrap intervals, using 10,000 resamples of seeds, nested months, and paired starts. Exact one-sided sign-flip tests \citep{ernst2004permutation} are Holm-adjusted \citep{holm1979multiple} across the prespecified family of 24 comparisons. Long-horizon analyses are descriptive secondary outcomes.

Practical importance is assessed by comparing each unrounded mean relative RMSE improvement with a threshold of 0.5\% for temperature and 5\% for height and precipitation. The threshold classification is based on the mean effect; bootstrap intervals describe uncertainty around that estimate.

For the three-seed training--inference pilot, Figure~\ref{fig:c3-completed-contrasts} reports descriptive Student-$t$ intervals and Table~\ref{tab:c3-factorial-contrasts} reports seed means and ranges. The direct-supervision experiment uses 20,000 paired cluster-bootstrap resamples of its nine adaptation seeds.

\subsection{Physical-Consistency Pilot}
\label{app:physical-pilot}

The physical-consistency comparison uses seeds 7, 11, and 23 and seven forecast starts, 1, 5, 9, 13, 17, 21, and 25 October 1940, with 72-hour rollouts. For each seed, dry-air drift is the maximum absolute change in area-weighted global-mean dry-air surface pressure over forecast starts and saved leads, relative to initialization. We then average these maxima across seeds. Water-budget residuals are area-weighted absolute means, averaged over starts, saved leads, and seeds. The negative-precipitation fraction is area-weighted and averaged over starts, the 24- and 72-hour endpoints, and seeds.

\section{Supplementary ACE2 Results}

\subsection{Matched-Control Comparisons}
\label{app:matched-results}

\begin{figure}[!htbp]
\centering
\includegraphics[width=\linewidth]{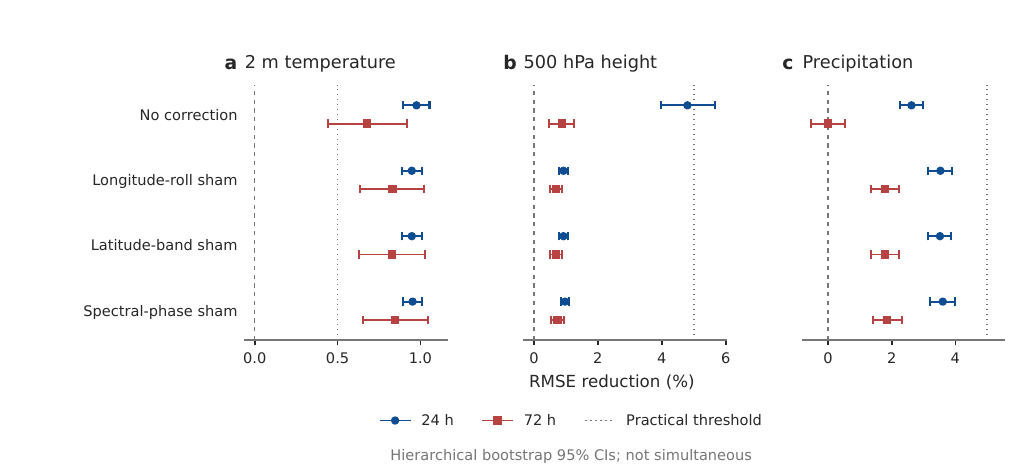}
\caption{\textbf{All fixed matched-condition comparisons.} Points show the official correction's relative RMSE reduction against each control; blue circles identify 24 h and red squares identify 72 h. Intervals are pointwise hierarchical-bootstrap 95\% confidence intervals. The dotted line is the variable-specific practical threshold and the dashed line denotes equal error. All 24 prespecified comparisons are shown. There are nine independent training seeds, with months and initial conditions nested within seeds. Separate numerical axes are retained for each variable.}
\label{fig:c2-primary-effects}
\end{figure}

Figure~\ref{fig:c2-primary-effects} displays all 24 matched-control effects with intervals; numerical results appear in Table~\ref{tab:c2-fixed-contrasts}.

\subsection{Long-Horizon Comparisons}

\begin{figure}[!htbp]
\centering
\includegraphics[width=\linewidth]{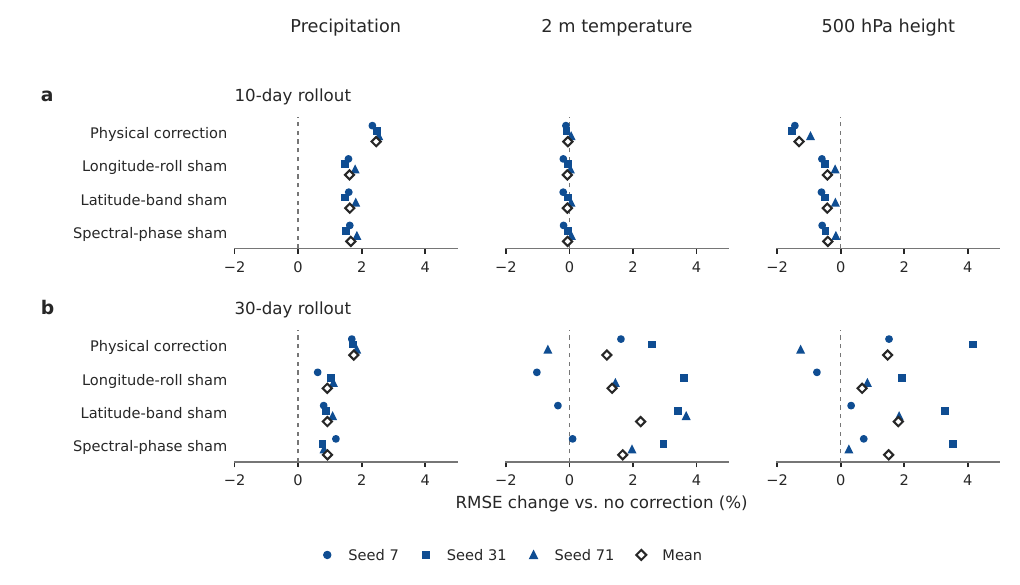}
\caption{\textbf{Long-horizon RMSE changes relative to no correction.} Rows show 10- and 30-day rollouts and columns show forecast variables. Blue circles, squares and triangles identify training seeds 7, 31 and 71; open black diamonds show their mean. The dashed line denotes no change. Negative values indicate lower RMSE; this sign convention differs from the RMSE-reduction figures. No confidence intervals are drawn.}
\label{fig:c2-long-horizon}
\end{figure}

The long-horizon matched-regime evaluation uses February, May, August, and November 1940 and training seeds 7, 31, and 71. Ten-day forecasts use four initialization dates per month; 30-day forecasts use one. Figure~\ref{fig:c2-long-horizon} reports window-averaged RMSE changes relative to the matched no-correction model.

\subsection{Training--Inference Comparisons}

\begin{table}[!htbp]
\caption{Descriptive effects of correction during adaptation and inference.
Each entry reports the mean on the first line and [minimum, maximum] on the second line across three training seeds.
Blocks 1--3 report relative RMSE reductions (\%); positive values indicate lower RMSE than the specified uncorrected reference.
Block 4 reports differences between relative reductions in percentage points.}
\label{tab:c3-factorial-contrasts}
\centering
\fontsize{9}{10}\selectfont
\setlength{\tabcolsep}{3pt}
\renewcommand{\arraystretch}{1.1}
\begin{tabular*}{\linewidth}{@{\extracolsep{\fill}}lrrrr@{}}
\toprule
\multicolumn{5}{@{}l@{}}{\textbf{1. Retaining correction at inference (\%)}} \\[2pt]
\multicolumn{5}{@{}l@{}}{\fontsize{8}{9}\selectfont Adapted with correction; reference: the same predictor run without correction.} \\[3pt]
Variable & 24 h & 72 h & 10 d & 30 d \\
\cmidrule(l){2-5}
2 m temperature
& \shortstack[r]{$-0.47$\\[1pt]{\fontsize{8}{9}\selectfont $[-0.50,\,-0.46]$}}
& \shortstack[r]{$+1.67$\\[1pt]{\fontsize{8}{9}\selectfont $[1.55,\,1.76]$}}
& \shortstack[r]{$+10.44$\\[1pt]{\fontsize{8}{9}\selectfont $[10.11,\,10.64]$}}
& \shortstack[r]{$+0.40$\\[1pt]{\fontsize{8}{9}\selectfont $[-4.54,\,5.58]$}} \\[4pt]
500 hPa height
& \shortstack[r]{$+51.58$\\[1pt]{\fontsize{8}{9}\selectfont $[50.61,\,52.37]$}}
& \shortstack[r]{$+20.55$\\[1pt]{\fontsize{8}{9}\selectfont $[19.11,\,21.94]$}}
& \shortstack[r]{$+9.03$\\[1pt]{\fontsize{8}{9}\selectfont $[8.80,\,9.17]$}}
& \shortstack[r]{$+1.70$\\[1pt]{\fontsize{8}{9}\selectfont $[-1.03,\,4.23]$}} \\[4pt]
Precipitation
& \shortstack[r]{$+38.04$\\[1pt]{\fontsize{8}{9}\selectfont $[37.96,\,38.10]$}}
& \shortstack[r]{$+7.75$\\[1pt]{\fontsize{8}{9}\selectfont $[7.64,\,7.83]$}}
& \shortstack[r]{$-15.35$\\[1pt]{\fontsize{8}{9}\selectfont $[-16.00,\,-14.47]$}}
& \shortstack[r]{$-20.13$\\[1pt]{\fontsize{8}{9}\selectfont $[-22.45,\,-17.21]$}} \\[4pt]
\addlinespace[6pt]
\midrule
\multicolumn{5}{@{}l@{}}{\textbf{2. Adaptation with correction (\%)}} \\[2pt]
\multicolumn{5}{@{}l@{}}{\fontsize{8}{9}\selectfont Both run without correction; reference: the predictor adapted without correction.} \\[3pt]
Variable & 24 h & 72 h & 10 d & 30 d \\
\cmidrule(l){2-5}
2 m temperature
& \shortstack[r]{$+1.34$\\[1pt]{\fontsize{8}{9}\selectfont $[1.33,\,1.36]$}}
& \shortstack[r]{$-1.69$\\[1pt]{\fontsize{8}{9}\selectfont $[-1.80,\,-1.61]$}}
& \shortstack[r]{$-11.51$\\[1pt]{\fontsize{8}{9}\selectfont $[-11.94,\,-11.22]$}}
& \shortstack[r]{$-3.24$\\[1pt]{\fontsize{8}{9}\selectfont $[-5.39,\,0.10]$}} \\[4pt]
500 hPa height
& \shortstack[r]{$-93.72$\\[1pt]{\fontsize{8}{9}\selectfont $[-96.76,\,-90.27]$}}
& \shortstack[r]{$-27.10$\\[1pt]{\fontsize{8}{9}\selectfont $[-29.21,\,-24.98]$}}
& \shortstack[r]{$-9.17$\\[1pt]{\fontsize{8}{9}\selectfont $[-10.13,\,-8.50]$}}
& \shortstack[r]{$-0.44$\\[1pt]{\fontsize{8}{9}\selectfont $[-2.72,\,0.80]$}} \\[4pt]
Precipitation
& \shortstack[r]{$-59.73$\\[1pt]{\fontsize{8}{9}\selectfont $[-59.82,\,-59.65]$}}
& \shortstack[r]{$-7.13$\\[1pt]{\fontsize{8}{9}\selectfont $[-7.17,\,-7.04]$}}
& \shortstack[r]{$+13.48$\\[1pt]{\fontsize{8}{9}\selectfont $[13.30,\,13.81]$}}
& \shortstack[r]{$+15.69$\\[1pt]{\fontsize{8}{9}\selectfont $[14.78,\,16.93]$}} \\[4pt]
\addlinespace[6pt]
\midrule
\multicolumn{5}{@{}l@{}}{\textbf{3. Adding correction at inference (\%)}} \\[2pt]
\multicolumn{5}{@{}l@{}}{\fontsize{8}{9}\selectfont Adapted without correction; reference: the same predictor run without correction.} \\[3pt]
Variable & 24 h & 72 h & 10 d & 30 d \\
\cmidrule(l){2-5}
2 m temperature
& \shortstack[r]{$-0.01$\\[1pt]{\fontsize{8}{9}\selectfont $[-0.01,\,-0.01]$}}
& \shortstack[r]{$+0.01$\\[1pt]{\fontsize{8}{9}\selectfont $[0.01,\,0.02]$}}
& \shortstack[r]{$+0.13$\\[1pt]{\fontsize{8}{9}\selectfont $[0.01,\,0.33]$}}
& \shortstack[r]{$+0.19$\\[1pt]{\fontsize{8}{9}\selectfont $[-3.76,\,3.41]$}} \\[4pt]
500 hPa height
& \shortstack[r]{$+0.11$\\[1pt]{\fontsize{8}{9}\selectfont $[0.03,\,0.16]$}}
& \shortstack[r]{$-0.12$\\[1pt]{\fontsize{8}{9}\selectfont $[-0.13,\,-0.10]$}}
& \shortstack[r]{$+0.19$\\[1pt]{\fontsize{8}{9}\selectfont $[-0.01,\,0.50]$}}
& \shortstack[r]{$+2.44$\\[1pt]{\fontsize{8}{9}\selectfont $[1.15,\,4.84]$}} \\[4pt]
Precipitation
& \shortstack[r]{$+0.23$\\[1pt]{\fontsize{8}{9}\selectfont $[0.22,\,0.24]$}}
& \shortstack[r]{$+1.16$\\[1pt]{\fontsize{8}{9}\selectfont $[1.04,\,1.25]$}}
& \shortstack[r]{$+1.21$\\[1pt]{\fontsize{8}{9}\selectfont $[1.12,\,1.31]$}}
& \shortstack[r]{$+3.12$\\[1pt]{\fontsize{8}{9}\selectfont $[1.96,\,5.02]$}} \\[4pt]
\addlinespace[6pt]
\midrule
\multicolumn{5}{@{}l@{}}{\textbf{4. Difference in correction effects (percentage points)}} \\[2pt]
\multicolumn{5}{@{}l@{}}{\fontsize{8}{9}\selectfont Block 1 minus block 3, paired within each seed; positive means a larger effect in block 1.} \\[3pt]
Variable & 24 h & 72 h & 10 d & 30 d \\
\cmidrule(l){2-5}
2 m temperature
& \shortstack[r]{$-0.47$\\[1pt]{\fontsize{8}{9}\selectfont $[-0.49,\,-0.45]$}}
& \shortstack[r]{$+1.66$\\[1pt]{\fontsize{8}{9}\selectfont $[1.54,\,1.74]$}}
& \shortstack[r]{$+10.32$\\[1pt]{\fontsize{8}{9}\selectfont $[10.07,\,10.57]$}}
& \shortstack[r]{$+0.20$\\[1pt]{\fontsize{8}{9}\selectfont $[-5.47,\,3.92]$}} \\[4pt]
500 hPa height
& \shortstack[r]{$+51.47$\\[1pt]{\fontsize{8}{9}\selectfont $[50.59,\,52.24]$}}
& \shortstack[r]{$+20.66$\\[1pt]{\fontsize{8}{9}\selectfont $[19.24,\,22.03]$}}
& \shortstack[r]{$+8.84$\\[1pt]{\fontsize{8}{9}\selectfont $[8.30,\,9.18]$}}
& \shortstack[r]{$-0.74$\\[1pt]{\fontsize{8}{9}\selectfont $[-5.87,\,2.88]$}} \\[4pt]
Precipitation
& \shortstack[r]{$+37.82$\\[1pt]{\fontsize{8}{9}\selectfont $[37.74,\,37.86]$}}
& \shortstack[r]{$+6.59$\\[1pt]{\fontsize{8}{9}\selectfont $[6.58,\,6.60]$}}
& \shortstack[r]{$-16.55$\\[1pt]{\fontsize{8}{9}\selectfont $[-17.12,\,-15.78]$}}
& \shortstack[r]{$-23.25$\\[1pt]{\fontsize{8}{9}\selectfont $[-24.41,\,-22.23]$}} \\[4pt]
\bottomrule
\end{tabular*}
\par\vspace{4pt}
\begin{minipage}{\linewidth}
\fontsize{8}{9}\selectfont
\raggedright
Within each seed, RMSE is averaged over the same four evaluation months before relative effects are calculated.
Each inference effect uses its own predictor's uncorrected RMSE as the denominator.
Block 4 subtracts the two relative inference effects within each seed before computing the mean and range.

\end{minipage}
\end{table}

\subsection{Direct-Supervision Seed Comparisons}
\label{app:direct-supervision}

\begin{figure}[!htbp]
\centering
\includegraphics[width=\linewidth]{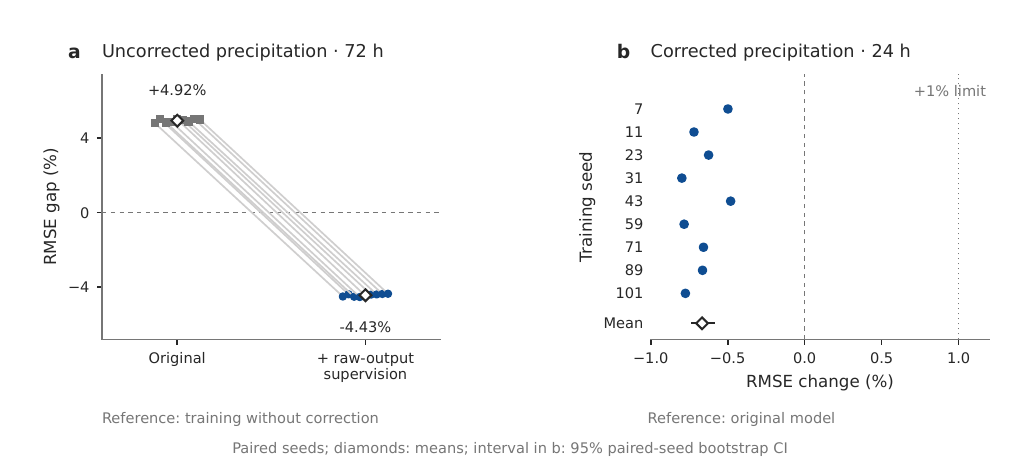}
\caption{\textbf{Direct supervision of uncorrected predictions.} (a) Paired 72-hour precipitation RMSE gaps relative to training without correction. Gray squares and blue circles show models before and after raw-output supervision; lines connect the same seed and open diamonds mark means. (b) Corrected 24-hour precipitation RMSE changes relative to the original correction-trained model. Circles show nine seeds; the mean diamond and horizontal interval show the mean and its 95\% paired-seed bootstrap interval. Dashed lines mark zero; the dotted +1\% line is the prespecified allowable increase. The reference model differs between panels a and b.}
\label{fig:raw-loss-decision}
\end{figure}

Figure~\ref{fig:raw-loss-decision} shows the individual paired-seed changes summarized in Table~\ref{tab:raw-loss-summary}.

\section{Additional Systems and Controls}

\subsection{Within-Family Energy-Correction Evaluation}

The ACE2S-SHiELD+ experiment evaluates total-energy correction using the model and released checkpoints of \citet{clark2026disentangling}, from two seed families, rs0 and rs1. Each checkpoint is evaluated from four forecast starts: 1 January, 1 April, 1 July, and 1 October 1979.

The fixed-checkpoint and unequal-budget between-checkpoint comparisons are distinguished in the main text. Both between-checkpoint forecasts disable energy correction at inference.

\begin{figure}[!htbp]
\centering
\includegraphics[width=\linewidth]{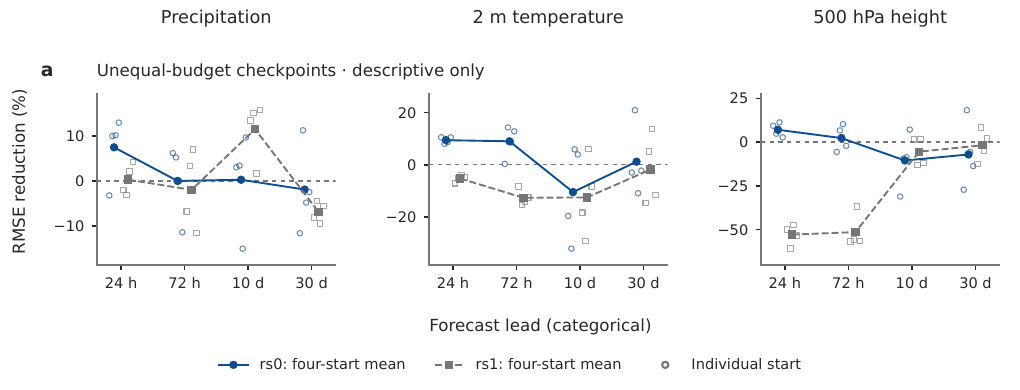}
\caption{\textbf{Descriptive comparison of unequal-budget ACE2S checkpoints.} Both checkpoint families are evaluated without the inference energy corrector. Positive RMSE reduction favors the checkpoint adapted with correction. Hollow marks show initial-condition-specific effects and connected marks show each seed's four-start mean. Circles with solid blue lines identify rs0; squares with dashed gray lines identify rs1. The four starts are January, April, July and October 1979, ordered left to right within each seed--lead group. Lead times are categorical. Unequal adaptation budgets confound the between-checkpoint contrast; it does not isolate a causal effect of correction during adaptation.}
\label{fig:ace2s-budget-context}
\end{figure}

Figure~\ref{fig:ace2s-replication} in the main text shows the fixed-checkpoint inference comparisons. Figure~\ref{fig:ace2s-budget-context} below reports the separate, budget-confounded between-checkpoint comparison.

\subsection{ConCerNet Evaluation Criteria}

The following seed-level criteria apply to the nine-seed ConCerNet experiment at its primary horizon, one model-time unit.

Let $E_{T,0}$ and $E_{U,0}$ denote errors without inference projection for models trained with and without projection, respectively. Let $E_{T,P}$ and $E_{T,S}$ denote errors for projection-trained models evaluated with physical and sham projection. The excess uncorrected error ratio is
\begin{equation}
Q_s
=
\frac{E_{T,0}-E_{U,0}}{E_{U,0}}.
\end{equation}
Adding one gives the total error ratio reported in the main text.

The physical-projection error reduction is
\begin{equation}
P_s
=
100\frac{E_{T,0}-E_{T,P}}{E_{T,0}},
\end{equation}
and its advantage over the sham projection is
\begin{equation}
S_s
=
100\frac{E_{T,S}-E_{T,P}}{E_{T,0}},
\end{equation}
reported in percentage points. These quantities are computed within each seed.

Invariant drift is the mean absolute change in the conserved state sum
$g(x)=\sum_j x_j$ from its initial value. Its reduction under physical projection is
\begin{equation}
100\frac{D_{T,0}-D_{T,P}}{D_{T,0}}.
\end{equation}

Table~\ref{tab:concernet-decision} reports the decision criteria and results.

\begin{table}[!htbp]
\caption{ConCerNet decision criteria and results at one simulation-time unit.}
\label{tab:concernet-decision}
\centering
\small
\setlength{\tabcolsep}{4pt}
\renewcommand{\arraystretch}{1.15}
\begin{tabular}{@{}p{0.41\linewidth}lrrl@{}}
\toprule
Comparison & Unit & \shortstack{Mean/\\count} & \shortstack{Lower 95\%\\CI endpoint} & Criterion \\
\midrule
Excess uncorrected vector-field error: projection-trained vs.\ trained without projection
& ratio & 25.874 & 15.516 & $\bar{x}>0,\ L>0$ \\
Vector-field error reduction: physical projection vs.\ no projection
& \% & 92.65 & 86.53 & $\bar{x}>0,\ L>0$ \\
Error-reduction difference: physical vs.\ sham projection
& pp & 95.47 & 87.97 & $\bar{x}>0,\ L>0$ \\
Invariant-drift reduction: physical vs.\ no projection
& \% & 100.00 & --- & $\bar{x}\geq90$ \\
Seeds with nonnegative excess uncorrected error
& seeds & 9/9 & --- & $\geq8/9$ \\
\bottomrule
\end{tabular}
\par\smallskip
\begin{minipage}{\linewidth}
\footnotesize
Values are seed means ($\bar{x}$), except the final count.
$L$ is the lower endpoint of a two-sided 95\% percentile bootstrap interval for the seed mean, based on 20,000 resamples of nine seeds.
The dash denotes a criterion that does not use an interval.
All prediction errors are trajectory-based vector-field RMSE.
The excess ratio equals the total error ratio minus one.
The physical--sham contrast is measured in percentage points.
The invariant-drift reduction of 100.00\% is rounded.
\end{minipage}
\end{table}

\end{document}